\documentclass[12pt,english,floatfix,nofootinbib,superscriptaddress,aps,prd,preprint]{revtex4}
\usepackage[utf8]{inputenc}
\usepackage{float}
\usepackage{array}
\usepackage{lipsum}
\usepackage{graphicx}
\usepackage{amsmath,amsthm,amsfonts,amssymb}
\usepackage{graphicx}
\usepackage[english]{babel}
\usepackage{color}
\usepackage{subfig}
\usepackage{caption}
\usepackage{tensor}
\usepackage{esint}
\usepackage[dvips]{epsfig}
\usepackage[dvips]{graphicx}
\usepackage{float}
\usepackage{units}
\usepackage{textcomp}
\usepackage{mathrsfs}
\usepackage{amsmath}
\usepackage[makeroom]{cancel}
\usepackage{amssymb}
\usepackage{amsbsy}
\usepackage{amsfonts}
\usepackage{amssymb,mathrsfs,xcolor}
\usepackage{esint}
\usepackage{array}
\usepackage{graphicx}

\usepackage{hyperref}
\hypersetup{
    colorlinks,
    citecolor=blue,
    filecolor=green,
    linkcolor=purple,
    urlcolor=red,
}

\usepackage{slashed}

\newcommand{\ie}{\begin{equation}}
\newcommand{\fe}{\end{equation}}
\newcommand{\se}{\begin{eqnarray}}
\newcommand{\ff}{\end{eqnarray}}

\usepackage{hyperref}
\hypersetup{colorlinks,breaklinks,
			citecolor=[rgb]{0,0.0,1.0},
            urlcolor=[rgb]{0.0,0.0,1.0},
            linkcolor=[rgb]{0,0.5,0.9}}

\begin{document}

\title{Vacuum solution within a metric-affine bumblebee gravity}

\author{A. A. Ara\'{u}jo Filho}
\email{dilto@fisica.ufc.br}

\affiliation{Departamento de Física, Universidade Federal da Paraíba, Caixa Postal 5008, 58051-970, João Pessoa, Paraíba,  Brazil.}
\affiliation{Departamento de Física Teórica and IFIC,
Centro Mixto Universidad de Valencia--CSIC. Universidad
de Valencia, Burjassot-46100, Valencia, Spain.}

\author{J. R. Nascimento}
\email{jroberto@fisica.ufpb.br}
\affiliation{Departamento de Física, Universidade Federal da Paraíba, Caixa Postal 5008, 58051-970, João Pessoa, Paraíba,  Brazil.}

\author{A. Yu. Petrov}
\email{petrov@fisica.ufpb.br}
\affiliation{Departamento de Física, Universidade Federal da Paraíba, Caixa Postal 5008, 58051-970, João Pessoa, Paraíba,  Brazil.}

\author{P. J. Porfírio}
\email{pporfirio@fisica.ufpb.br}
\affiliation{Departamento de Física, Universidade Federal da Paraíba, Caixa Postal 5008, 58051-970, João Pessoa, Paraíba,  Brazil.}




\date{\today}

\begin{abstract}

We consider a metric--affine extension to the gravitational sector of the Standard--Model Extension for the Lorentz--violating coefficients $u$ and $s^{\mu\nu}$. The general results, which are applied to a specific model called metric--affine bumblebee gravity, are obtained. A Schwarzschild--like solution, incorporating effects of the Lorentz symmetry breaking through coefficient $X=\xi b^2$, is found. Furthermore, a complete study of the geodesics trajectories of particles has been accomplished in this background, emphasizing the departure from general relativity. We also compute the advance of Mercury's perihelion and the deflection of light within the context of the weak field approximation, and we verify that there exist two new contributions ascribed to the Lorentz symmetry breaking. As a phenomenological application, we compare our theoretical results with observational data in order to estimate the coefficient $X$.

\end{abstract}


\maketitle


\section{Introduction}
\label{sec:intro}

The problem of a consistent implementation of Lorentz symmetry breaking (LSB) within the gravitational scenario differs crucially from the construction of Lorentz-breaking extensions for non-gravitational field theories. Flat spacetimes admit Lorentz--breaking additive terms, such as, Carroll--Field--Jackiw \cite{CFJ}, aether time \cite{aether} and other ones (see f.e. \cite{ColKost}), and they can fundamentally be constructed on the basis of a constant vector (tensor) contracted to some functions of fields and its derivatives. On the other hand, in curved spacetimes, such features cannot properly be applied. 

Indeed, constant tensors being well-defined in Minkowski spacetime, for instance, the simple conditions like $\partial_{\mu}k_{\nu}=0$ are not able to be introduced in an analogous manner in curved spacetimes. The term $\partial_{\mu}k_{\nu}=0$ is clearly noncovariant, and its natural covariant extension, namely, $\nabla_{\mu}k_{\nu}=0$, entails severe restrictions to the spacetime geometries (the so--called no--go constraints \cite{KosLi}). 
As a result, it turns out that the most natural way to incorporate (local) Lorentz violation into gravitational theories is based on the mechanism of spontaneous symmetry breaking. In this case, Lorentz/CPT violating (LV) coefficients (operators) arise as vacuum expectation values (VEV) of dynamical tensor fields, which are driven by nontrivial potentials.

The generic effective field framework, describing all possible coefficients for Lorentz/CPT violation, is the well--known Standard--Model Extension (SME) \cite{KosGra}. In particular, its gravitational sector has been defined in a Riemann--Cartan manifold, in which the torsion is treated as a dynamic geometrical quantity besides the metric. Although the gravity SME sector is defined in a non-Riemannian background, up to now, most studies 
have been performed within the metric approach of gravity -- where the metric is the only dynamical geometric field. In this context, such investigations are mainly based on obtaining some exact solutions to different models, accommodating LSB in curved spacetimes, e.g., bumblebee gravity \cite{Bertolami,Casana,Santos1,Santos2,Santos3, Maluf, Maluf2, Kumar, Xu:2022frb}, Einstein-Aether model \cite{Ted}, parity--violating models \cite{Jackiw, Mir, Bartolo, Conroy, Rao}, and Chern--Simons modified gravity \cite{PP1,PP2,PP3}. Signals for Lorentz violation within the pseudo-Riemannian approach in Solar system experiments also were tested \cite{Bailey1, Bailey2, Bailey3}.

On the other hand, in the literature, despite having the vast majority of works concerning modified theories of gravity in the usual metric approach, it is interesting to take into account more generic geometrical frameworks. Among more specific motivations for exploiting theories of gravity in a Riemann-Cartan background, we can point out the induction of gravitational topological term \cite{Paulo}. Another interesting non-Riemannian geometry that has been considered in the literature is the Finsler one \cite{Bao}, which possesses a variety of works linking such a geometry to LSB in recent years \cite{Foster, KosE, Sch1, CollM, Sch2}.

However, the most compelling  generalization of the metric approach is the so--called metric--affine (Palatini) formalism, in which metric and connection are supposed to be independent  dynamic geometrical quantities (for a discussion and some interesting results within Palatini approach, see f.e. \cite{Ghil1,Ghil2}, and references therein). In this scenario, LSB still remains no much explored in the literature, even though there exist some recent works involving bumblebee gravity  \cite{Paulo2, Paulo3, Paulo4}. In particular, the authors have found the field equations and solved them; additionally, they investigated the stability conditions and the associated dispersion relations for different matter sources in the weak field and post-Newtonian limit. Apart from that, at the quantum level, they have also computed the divergent piece of one-loop corrections to the spinor effective action in two different ways: using the diagrammatic method (in this case, disregarding the gravitational effects) and using the Barvinsky-Vilkovisky technique (considering the gravitational effects). Along the
same line, a metric-affine version of the Chern--Simons modified gravity invariant under projective transformations has been proposed \cite{Paulo5, Boudet1, Boudet2}. The authors solved the field equations adopting a perturbative scheme since the exact solution of the connection equation remains unknown, and the quasinormal modes of Schwarzschild black holes were carried out in this model.

In this work, we deal with the gravitational sector of the SME in the metric--affine approach. We shall assume LV coefficient $t^{\mu\nu\alpha\beta}=0$, while the other ones, namely, $u$ and $s^{\mu\nu}$, are nonzero coefficients. In particular, we find the first exact solution for a particular metric--affine bumblebee gravity model (different from those proposed in \cite{Paulo2, Paulo3, Paulo4}), filling this gap in the literature. Furthermore, we investigate the role played by the LSB coefficients, confronting our theoretical results with the observational data of the deflection of light. Also, we calculate the advance of Mercury's perihelion.

This paper is organized as follows. In section \ref{general}, we propose a metric--affine generalization to the gravitational sector of the SME, setting the LV coefficient  $t^{\mu\nu\alpha\beta}=0$; we find the field equations disregarding fermions as matter sources. In section \ref{traceless}, we take a particular model called traceless metric--affine bumblebee gravity, considering LV coefficient $s^{\mu\nu}$ to be traceless. We derive and solve the field equations, using the general expressions found in section \ref{general}. Additionally, in section \ref{Application}, we obtain an exact solution, which describes a static and spherically symmetric spacetime. Next, we present a discussion to the geodesic trajectories of particles as well as the consequences of LSB on them. Also, we provide the bounds to LSB coefficients, from astrophysics observational data of the deflection of light and we compute the advance of Mercury's perihelion. Finally, in section \ref{summary}, we write the conclusions to this manuscript.

In this paper, we use the following conventions: the metric signature $(-,+,+,+)$, $\kappa^2=8\pi G$ and the Riemann tensor is defined by $R^{\mu}_{\,\,\,\nu\alpha\beta}=\partial_{\alpha}\Gamma^{\mu}_{\,\,\,\beta\nu}+\Gamma^{\mu}_{\,\,\,\alpha\lambda}\Gamma^{\lambda}_{\,\,\,\beta\nu}-\left(\alpha\longleftrightarrow\beta\right)$.


\section{The general setup}
\label{general}

We start this section by presenting a metric--affine generalization of the gravitational sector of the SME \cite{KosGra}. Similarly to the metric case, the action of this sector can be cast into the following form
\begin{eqnarray}
    \nonumber\mathcal{S} &=& \frac{1}{2 \kappa^{2}} \int \mathrm{d}^{4}x \sqrt{-g} \left\{ (1-u)R(\Gamma) + s^{\mu\nu}R_{\mu\nu}(\Gamma) + t^{\mu\nu\alpha\beta}R_{\mu\nu\alpha\beta}(\Gamma)    \right\}+\mathcal{S}_{mat}(g_{\mu\nu},\psi)+\\
    &+& \mathcal{S}_{coe}(g_{\mu\nu},u,s^{\mu\nu},t^{\mu\nu\alpha\beta}),
    \label{1}
\end{eqnarray}
where the geometrical quantities $R(\Gamma)\equiv g^{\mu\nu}R_{\mu\nu}(\Gamma)$, $R_{\mu\nu}(\Gamma)$, and $R^{\mu}_{\,\,\,\nu\alpha\beta}(\Gamma)$ are the Ricci scalar, Ricci tensor and Riemann tensor respectively, and $\mathcal{S}_{mat}$ is the action describing the contributions of the matter sources, which are supposed to be coupled to the metric only\footnote{Note that fermions possess a natural coupling to the connection; therefore, based on our assumption, we are disregarding spinors (for the sake of convenience) and we just consider the bosonic matter sources minimally coupled to the metric.}. As it was pointed out before, the action is defined in the metric--affine (Palatini) formalism, in which metric and connection are taken to be independent dynamical quantities {\it a priori}. Furthermore, $u=u(x)$, $s^{\mu\nu}=s^{\mu\nu}(x)$ and $t^{\mu\nu\alpha\beta}=t^{\mu\nu\alpha\beta}(x)$ are coefficients (fields) responsible for the explicit (local) Lorentz symmetry breaking, as exhaustively discussed in \cite{KosGra}. It is worth to be mentioned that the background field $s^{\mu\nu}$ exhibits the same symmetries of the Ricci tensor. Nevertheless, in the present work, let us assume that it is a symmetric second-rank tensor, $s^{\mu\nu}=s^{(\mu\nu)}$. Thereby, it only couples to the symmetric piece of the Ricci tensor. In addition, $t^{\mu\nu\alpha\beta}$ possesses the same symmetries of the Riemann tensor. Finally, the last term, in Eq.(\ref{1}), i.e., $\mathcal{S}_{coe}$, accounts for the dynamical contributions of the Lorentz-violating coefficients.

In this work, we shall concentrate our efforts on the nontrivial effects of Lorentz symmetry breaking involving both the Ricci tensor and  scalar, so that we can restrict the coefficients $s^{\mu\nu}$ and $u$ to be nonzero, whilst $t^{\mu\nu\alpha\beta}$ is set to be zero. This happens in part because of the connection equation cannot be solved into a simple metric redefinition for a nontrivial $t^{\mu\nu\alpha\beta}$ parameter, i.e., this problem is known as ``{\it t}-puzzle'' \cite{Bonder:2015maa}. In this way, the action that we are interested in reads:
\begin{equation}
\mathcal{S} = \frac{1}{2 \kappa^{2}} \int \mathrm{d}^{4}x \sqrt{-g} \left\{ (1-u)R(\Gamma) + s^{\mu\nu}R_{\mu\nu}(\Gamma)    \right\}+\mathcal{S}_{mat}+\mathcal{S}_{coe}.
\label{S2}
\end{equation}
It is worth stressing out that the above action is invariant under projective transformations of the connection,
\begin{equation}
\Gamma^{\mu}_{\nu\alpha}\longrightarrow \Gamma^{\mu}_{\nu\alpha}+\delta^{\mu}_{\alpha}A_{\nu},
\label{Proj}
\end{equation}
where $A_{\alpha}$ is an arbitrary vector. It is easy to check that the Riemann tensor under the projective transformation, Eq.(\ref{Proj}), changes as follows:
\begin{equation}
    R^{\mu}_{\,\,\,\nu\alpha\beta}\longrightarrow R^{\mu}_{\,\,\,\nu\alpha\beta}-2\delta^{\mu}_{\nu}\partial_{[\alpha}A_{\beta]},
\end{equation}
as a consequence, the symmetric portion of the Ricci tensor is invariant under Eq.(\ref{Proj}), as well as, the whole action (\ref{S2}).

The model given by the action (\ref{S2}) belongs to a more generic class of gravitational theories called Ricci--based ones \cite{Afonso:2017bxr, BeltranJimenez:2017doy, Delhom:2021bvq}. It has been shown that, for this class of models, the projective invariance avoids the emergence of gravitational ghost--like propagating degrees of freedom \cite{AD}. We note that a bootstrap procedure, using the Palatini approach for spontaneous Lorentz symmetry breaking in gravity, is presented by the cardinal gravity model in \cite{Kostelecky:2009zr}, however, we follow a different procedure.


\subsection{Field equations}
\label{FEI}


\subsubsection{The connection equation}



Here, we develop the connection equation. To do so, we vary Eq. (\ref{S2}) with respect to the connection; then, we find the following field equation: 
\begin{equation}
\nabla^{(\Gamma)}_{\alpha}  \left[ \sqrt{-h} h^{\mu\nu}  \right] = \sqrt{-h}\left[T^{\mu}_{\,\,\alpha\lambda}h^{\nu\lambda}+T^{\lambda}_{\,\,\lambda\alpha}h^{\mu\nu}-\frac{1}{3}T^{\lambda}_{\,\,\lambda\beta}h^{\nu\beta}\delta^{\mu}_{\alpha}\right], \label{fieldequation}
\end{equation}
where $T^{\mu}_{\,\,\alpha\lambda}=2\Gamma^{\mu}_{\,\,[\alpha\lambda]}$ is the torsion tensor. We have also defined auxiliary metric 
\begin{equation}
h^{\mu\nu} \equiv\frac{\sqrt{-g}}{\sqrt{-h}}\left[\left(1-u\right)g^{\mu\nu} + s^{\mu\nu}\right].
\label{hg}
\end{equation}
For our purposes, since we will not take into account fermions, the torsional terms on the {\it r.h.s} of Eq.(\ref{fieldequation}) can be gauged away by means of an appropriate gauge--fixing (projective) choice as a result of the projective invariance of the model \cite{BeltranJimenez:2017doy}. Therefore, the solution of Eq.(\ref{fieldequation}) (up to an irrelevant projective mode) is given by
\begin{equation}
\Gamma\indices{^\mu_\nu_\alpha} = \left\{ \indices{^\mu_\nu_\alpha} \right\}^{(h)} = \frac{1}{2}h^{\mu\lambda}\left(-\partial_{\lambda}h_{\nu\alpha}+\partial_{\nu}h_{\alpha\lambda}+\partial_{\alpha}h_{\lambda\nu}\right),
\label{conn}
\end{equation}
which $\Gamma\indices{^\mu_\nu_\alpha}$ is the Levi-Civita connection of the $h_{\mu\nu}$ metric, and $h^{\mu\nu}$ is identified to be the inverse of $h_{\mu\nu}$. The next step is providing a relationship between $h_{\mu\nu}$ and $g_{\mu\nu}$. In order to do it, let us rewrite Eq.(\ref{hg}) in matrix form, namely,
\begin{equation}
    \sqrt{-h}\hat{h}^{-1}=\sqrt{-g}\hat{g}^{-1}\left[\left(1-u\right)\hat{I}+\hat{s}\right],
\end{equation}
where all terms that carry a hat are identified as being matrices. For example, $\hat{s}$ denotes the matrix form of $s^{\mu}_{\alpha}$ and $\hat{h}^{-1}$ denotes the matrix form of $h^{\mu\nu}$. Now, taking the determinant of the former equation, we are able to find
\begin{equation}
    \sqrt{-h}=\sqrt{-g}\sqrt{\mathrm{\det} \left[\left(1-u\right)\hat{I}+\hat{s}\right]}.
\label{dethg}
\end{equation}
Substituting Eq.(\ref{dethg}) in Eq.(\ref{hg}), one finds
\begin{equation}
    h^{\mu\nu}=\frac{1}{\sqrt{\mathrm{\det\hat{\Omega}^{-1}}}}g^{\mu\alpha}\left(\Omega^{-1}\right)^{\nu}_{\alpha},
\end{equation}
where we have defined the inverse of the deformation matrix by $\hat{\Omega}^{-1}\equiv \left(1-u\right)\hat{I}+\hat{s}$ and  
$\left(\Omega^{-1}\right)^{\nu}_{\alpha}\equiv\left(1-u\right)\delta^{\nu}_{\alpha}+s^{\nu}_{\alpha}$.
By means of a straightforward calculation, it turns out that
\begin{equation}
    h_{\mu\nu}=\sqrt{\mathrm{\det\hat{\Omega}^{-1}}}g_{\mu\lambda}\Omega^{\lambda}_{\nu}.
    \label{hk}
\end{equation}
Formally, the determinant is evaluated below 
\begin{equation}
    \mathrm{\det\hat{\Omega}^{-1}}=\mathrm{\det}\left[\left(1-u\right)\hat{I}+\hat{s}\right]=e^{\mbox{Tr}\ln{\left[\left(1-u\right)\hat{I}+\hat{s}\right]}}.
\end{equation}
This equation, in general, does not present an analytical closed form for generic $\hat{s}$ and $\hat{u}$. However, in some specific cases, we have such a feature. For instance, if we assume $s^{\mu}_{\nu}=\xi b^{\mu}b_{\nu}$ and $u=0$, where $\xi$ is a coupling constant and $b^{\mu}$ is a vector background field \cite{KosGra}. In this scenario, one can find an analytical expression for the determinant, namely,
\begin{equation}
    \mathrm{\det\hat{\Omega}^{-1}}=\mathrm{\det}\left(\hat{I}+\xi\hat{bb}\right)=e^{\mbox{Tr}\ln{\left(\hat{I}+\xi\hat{bb}\right)}}=1+\xi b^2,
\end{equation}
where $b^2 =b^{\mu}b_{\mu}$.

In order to ensure that the metric $h^{\mu\nu}$ is the inverse of $h_{\mu\nu}$, the deformation matrix must satisfy the following relation
\begin{equation}
    \delta^{\beta}_{\nu}=\left(1-u\right)\Omega^{\beta}_{\nu}+s^{\beta}_{\lambda}\Omega^{\lambda}_{\nu},
\end{equation}
or, in matrix form
\begin{equation}
    \hat{I}=\left(1-u\right)\hat{\Omega}+\hat{s}\cdot\hat{\Omega}.
    \label{Io}
\end{equation}
Such an equation tells us that $\hat{\Omega}$ should be a function of $\hat{s}$ and $\hat{u}$, and it can entirely be determined once the explicit form of $\hat{s}$ and $\hat{u}$ are known. For instance, assuming that the coefficient for Lorentz violation $s^{\mu\nu}$ is traceless, $s\indices{^\mu_\mu}=0$,  it entails that the most general $s^{\mu\nu}$ possesses nine independents degrees of freedom which, in turn, determine the structure of $\hat{\Omega}$ ruled by Eq.(\ref{Io}). We shall give an explicit example of how to calculate the deformation metric from the LV coefficients in the next section.

The action Eq.(\ref{S2}) admits an Einstein frame (tilded frame) form. To see that, we substitute Eq.(\ref{hg}) into Eq.(\ref{S2}),
\begin{equation}
    \tilde{\mathcal{S}} = \frac{1}{2 \kappa^{2}} \int \mathrm{d}^{4}x \sqrt{-h} R(h) + \mathcal{\tilde{S}}_{mat}(h_{\mu\nu}, u, s_{\mu\nu}, \psi)+\mathcal{\tilde{S}}_{coe}(h_{\mu\nu}, u, s_{\mu\nu}),
    \label{ck}
\end{equation}
where $R(h)\equiv h^{\mu\nu}R_{\mu\nu}(h)$.

A remarkable feature of the Einstein frame representation that should be emphasized is that the LV coefficients are moved from the gravitational to the matter sector. We shall see that this is the key to solving the vacuum field equations. Note further that  the Lorentz symmetry breaking model in Eq.(\ref{ck}) can be 
viewed as an Einstein--Hilbert action of $h_{\mu\nu}$ coupled to new bosonic matter sources, interacting
with Lorentz symmetry breaking coefficients. As a consequence, the dynamical field equations for the metric $h_{\mu\nu}$
must be Einstein-like ones incremented by non-linear couplings of the matter sources, $s^{\mu\nu}$ and $u$.



\subsubsection{The metric equation}

Here, to continue our study, we present the metric equation. In this sense, we have to vary Eq.(\ref{S2}) with respect to the metric; after that, we obtain the following expression:
\begin{equation}
 \left(1-u\right)R_{(\mu\nu)}(\Gamma) - \frac{1}{2} g_{\mu\nu}\left[ \left(1-u\right)R(\Gamma) + s^{\alpha\beta}R_{\alpha\beta}(\Gamma)\right] + 2 s^{\beta}  \!\,_{(\mu}R(\Gamma)_{\nu)\beta}   = \kappa^{2} T_{\mu\nu}, \label{equationofmotion1}
\end{equation}
where the stress-energy tensor may be split into two pieces:
\begin{equation}
    T_{\mu\nu}=T_{\mu\nu}^{(mat)}+T_{\mu\nu}^{(coe)},
\end{equation}
where
\begin{equation}
T_{\mu\nu}^{(mat)} = -\frac{2}{\sqrt{-g}}\frac{\delta(\sqrt{-g}\mathcal{L}_{mat})}{\delta g^{\mu\nu}}.
\label{matt}
\end{equation}
is the stress--energy tensor of the matter source contributions and  $T_{\mu\nu}^{(coe)}$ is the stress-energy tensor of the dynamical Lorentz-violating coefficients.

To exploit some properties of this model, let us first contract Eq.(\ref{equationofmotion1}) with $g^{\mu\nu}$ to get
\begin{equation}
\left(1-u\right)R(\Gamma) = - \kappa^{2} T,
\label{RT1}
\end{equation}
where, $T \equiv g^{\mu\nu}T_{\mu\nu}$. It turns out that the connection is algebraically related to the trace of the stress--energy tensor. Similarly, we contract Eq.(\ref{equationofmotion1}) with $s^{\mu\nu}$ to find 

\begin{equation}
s^{\mu\nu}R_{\nu\alpha}(\Gamma) \left[  \delta\indices{^\alpha_\mu}  \left( 1-u- \frac{1}{2}s \right) + 2 s\indices{^\alpha_\mu}          \right] = \kappa^{2} \left[ T^{(s)} - \frac{1}{2}s T \right], \label{scontraction}
\end{equation}
where $s \equiv s^{\mu\nu}g_{\mu\nu}$, and $T^{(s)} \equiv s^{\mu\nu}T_{\mu\nu}$. Plugging  Eqs.(\ref{RT1}) and (\ref{scontraction}) into Eq.(\ref{equationofmotion1}), we get
\begin{equation}
\begin{split}
   &( 1-u) R_{(\mu\nu)}(\Gamma)+2s\indices{^\alpha_{(\mu}}R_{\nu)\alpha}(\Gamma)+\frac{2}{2-s-2u}g_{\mu\nu}s\indices{^\beta_\alpha}s^{\alpha\lambda}R_{\lambda\beta}(\Gamma)=\kappa^{2}\bigg[T_{\mu\nu}-\\
   &-\frac{1}{2}g_{\mu\nu}T+\frac{g_{\mu\nu}}{2-s-2u}\left(T^{(s)}-\frac{1}{2}sT\right)\bigg].
    \label{re}
    \end{split}
\end{equation}
The {\it l.h.s} of Eq.(\ref{re}) shows the nonminimal interaction terms between $u$, $s^{\mu\nu}$, $R$ and $R_{\mu\nu}$. On the other hand, the {\it r.h.s} of Eq.(\ref{re}) depends only on $T_{\mu\nu}$ and $s^{\mu\nu}$. Furthermore, since the deformation matrix $\hat{\Omega}$ is known, Eq.(\ref{re}) may completely be rewritten in terms of $h_{\mu\nu}$, $s_{\mu\nu}$ and $T_{\mu\nu}$. In addition, the LV coefficients also satisfy their own field equations, and since this is not important for our purposes right now, we shall not explicit them.

Let us now explore the behavior of the theory in the first order limit to LV coefficients. With this, $\sqrt{\mathrm{\det\hat{\Omega}^{-1}}}\approx  \left(1+\dfrac{1}{2}s-2u\right)$, we have
\begin{equation}
\begin{split}
     h^{\mu\nu}&\approx \left(1+u-\frac{1}{2}s\right)g^{\mu\nu}+s^{\mu\nu},\\
     h_{\mu\nu}&\approx \left(1-u+\frac{1}{2}s\right)g_{\mu\nu}-s_{\mu\nu}.
     \end{split}
\end{equation}
Taking this limit into account, the action (\ref{ck}) reduces to
\begin{equation}
    \mathcal{S}=\frac{1}{2\kappa^2}\int d^{4}x\sqrt{-g}\left[\left(1-u\right)R(g)+s^{\mu\nu}R_{\mu\nu}\right]+...
\end{equation}
where  we got rid of the boundary terms, while ellipses mean high--order terms in $u$ and $s^{\mu\nu}$. It implies from the former equation that the action that was expanded, up to first-order in the LV coefficients is identical to the gravitational sector action of the SME defined in the metric approach. Such a result has been discussed in \cite{Bonder:2015maa, Bonder2}. However, none of them have addressed any possible exact solutions. We shall treat this issue next.

\section{The traceless metric-affine bumblebee model}
\label{traceless}

In this section, we shall focus on a particular metric--affine model known as the metric--affine bumblebee model. Such a case is achieved by taking the LV coefficients $u=0$ and $s^{\mu\nu}$ with the following form: 
\begin{equation}
\begin{split}
  s^{\mu\nu}=\xi\left(B^{\mu}B^{\nu}-\dfrac{1}{4}B^{2}g^{\mu\nu}\right), 
  \label{tracel}
  \end{split}
  \end{equation}
  where $B_{\mu}$ is the bumblebee field with $B^{2}\equiv g^{\mu\nu}B_{\mu}B_{\nu}$ and $s^{\mu\nu}$ is traceless that differs from that metric-affine bumblebee model considered in \cite{Paulo2,Paulo3,Paulo4}. The complete action for this particular bumblebee model takes the form
  \begin{eqnarray}
      \mathcal{S}_{B}&=&\int d^{4}x\,\sqrt{-g}\left[\frac{1}{2\kappa^2}\left(R(\Gamma)+\xi\left(B^{\mu}B^{\nu}-\frac{1}{4}B^{2}g^{\mu\nu}\right)R_{\mu\nu}(\Gamma)\right)-\frac{1}{4}B^{\mu\nu}B_{\mu\nu}-\right.\nonumber\\ &-&\left.V(B^{\mu}B_{\mu}\pm b^{2})\right]+
      \int d^{4}x\sqrt{-g}\mathcal{L}_{mat}(g_{\mu\nu},\psi),
      \label{bumb}
  \end{eqnarray}
  where $V(B^{\mu}B_{\mu}\pm b^2)$ is the potential that accounts for the spontaneous Lorentz symmetry breaking. In other words, the 
  bumblebee field acquires a nontrivial vacuum expectation value (VEV), when $<B_{\mu}>=b_{\mu}$, i.e., $b_{\mu}$ corresponds to a minimum of the potential. Furthermore, the field strength of $B_{\mu}$ is defined by
  \begin{equation}
      B_{\mu\nu}=(dB)_{\mu\nu},
  \end{equation}
 where $d$ means the exterior derivative operator. Note additionally that $b^{2}\equiv g^{\mu\nu}b_{\mu}b_{\nu}$ is chosen to be a real constant, so that $b_{\mu}$ is a norm--fixed vector.
 
 The action displayed in Eq. (\ref{bumb}) can be rewritten in a more convenient way as shown below
 \begin{equation}
 \begin{split}
     \mathcal{S}_{B}&=\int d^{4}x\,\sqrt{-g}\bigg\{\frac{1}{2\kappa^2}\left[\left(1-u\right)R(\Gamma)+s^{\mu\nu}R_{\mu\nu}(\Gamma)\right]-\frac{1}{4}B^{\mu\nu}B_{\mu\nu}-\\
     &-V(B^{\mu}B_{\mu}\pm b^{2})\bigg\}+\int d^{4}x\sqrt{-g}\mathcal{L}_{mat}(g_{\mu\nu},\psi),
     \end{split}
 \end{equation}
 where we define $u=\dfrac{\xi B^2}{4}$ and $s^{\mu\nu}=\xi B^{\mu}B^{\nu}$ in comparison to the gravitational sector of the SME. Notice that we have absorbed the traceless piece of $s^{\mu\nu}$ of Eq. (\ref{tracel}) into the coefficient $u$. Defining the matrices $\hat{u}=\dfrac{1}{4}B^{2}\hat{I}$ and  $\hat{s}=\xi\hat{BB}$, and implementing the general results outlined in the former section, for this particular case, we are able to find
\begin{equation}
    \hat{\Omega}^{-1}=\left(1-\frac{\xi B^2}{4}\right)\hat{I}+\xi\hat{BB}.
\end{equation}
Its determinant takes the following form
\begin{equation}
    \mathrm{\det\hat{\Omega}^{-1}}=\left(1-\frac{\xi B^2}{4}\right)^{4}\det\left(\hat{I}+\xi^{\prime} \hat{BB}\right),
\label{ww}
\end{equation}
where $\xi^{\prime}=\dfrac{\xi}{\left(1-\frac{\xi B^2}{4}\right)}$ must be understood as a function of $B$. Now, note that
\begin{equation}
 \det\left(\hat{I}+\xi^{\prime} \hat{BB}\right)=e^{\mbox{Tr}\ln{\left(\hat{I}+\xi^{\prime}\hat{BB}\right)}},
 \label{wi}
\end{equation}
which can analytically be solved by expanding the logarithm in power series 
\begin{equation}
    \begin{split}
    Tr\left(\xi^{\prime}\hat{BB}\right)&=\xi^{\prime}B^{\mu}B_{\mu}=\xi^{\prime}B^2;\\
    Tr\left[\left(\xi^{\prime}\hat{BB}\right)^{2}\right]&=(\xi^{\prime})^{2} B^{\mu}B_{\alpha}B^{\alpha}B_{\mu}=(\xi^{\prime})^{2}B^{4}=\left[Tr\left(\xi^{\prime}\hat{BB}\right)\right]^{2};\\
    &.\\
    &.\\
    &.\\
    Tr\left[\left(\xi^{\prime}\hat{BB}\right)^{n}\right]&=\left[Tr\left(\xi^{\prime}\hat{BB}\right)\right]^{n}.
    \end{split}
    \label{we}
\end{equation}
Therefore, combining Eqs.(\ref{ww}), (\ref{we}) and (\ref{wi}), we obtain
\begin{equation}
    \mathrm{\det\hat{\Omega}^{-1}}=\left(1-\frac{\xi B^2}{4}\right)^{3}\left(1+\frac{3}{4}\xi B^2\right).
\end{equation}
In possession of the above result, it is straightforward to get
\begin{eqnarray}
    h^{\mu\nu}=\frac{1}{\sqrt{\left(1-\frac{\xi B^2}{4}\right)\left(1+\frac{3}{4}\xi B^2\right)}}\left[g^{\mu\nu}+\frac{\xi B^{\mu}B^{\nu}}{\left(1-\frac{\xi B^2}{4}\right)}\right].
    \label{metric1}
\end{eqnarray}
In order to obtain $h_{\mu\nu}$, we need first to find the deformation matrix $\hat{\Omega}$ in an explicit manner. To do that, we note the fact that $\hat{\Omega}$ should be linear in $\hat{BB}$; thereby, its general form is
\begin{equation}
    \hat{\Omega}=A\hat{I}+C\hat{BB},
\end{equation}
where $A$ e $C$ are constants to be determined. Substituting it in Eq.(\ref{Io}), one finds 
\begin{equation}
    A=\frac{1}{\left(1-\frac{\xi B^2}{4}\right)},\,\,\, C=\frac{-\xi}{\left(1+\frac{3}{4}\xi B^2\right)\left(1-\frac{\xi B^2}{4}\right)}.
\end{equation}
As a result,
\begin{equation}
    \hat{\Omega}= \frac{1}{\left(1-\frac{\xi B^2}{4}\right)}\hat{I}-\frac{\xi}{\left(1-\frac{\xi B^2}{4}\right)\left(1+\frac{3}{4}\xi B^2\right)}\hat{BB}.
\end{equation}
Having found $\hat{\Omega}$, one straightforwardly concludes from Eq.(\ref{hk}) that
\begin{equation}
    h_{\mu\nu}=\sqrt{\left(1+\frac{3}{4}\xi B^2\right)\left(1-\frac{\xi B^2}{4}\right)}g_{\mu\nu}-\xi\sqrt{\frac{\left(1-\frac{\xi B^2}{4}\right)}{\left(1+\frac{3}{4}\xi B^2\right)}} B_{\mu}B_{\nu}.
    \label{metric2}
\end{equation}

We now deal with the metric equation presented in Eq. (\ref{equationofmotion1}). For our particular case of the bumblebee model, we have
\begin{equation}
\begin{split}
    &\left(1-\frac{\xi B^2}{4}\right)R_{(\mu\nu)}(\Gamma)-\frac{1}{2}g_{\mu\nu}R(\Gamma)+2\xi\left[B^{\alpha}B_{(\mu}R_{\nu)\alpha}(\Gamma)\right]-\frac{\xi}{4}B_{\mu}B_{\nu}R(\Gamma)-\\
    &-\frac{\xi}{2}g_{\mu\nu}B^{\alpha}B^{\beta}R_{\alpha\beta}(\Gamma)+
    \frac{\xi}{8}B^{2}g_{\mu\nu}R(\Gamma)=\kappa^2 T_{\mu\nu},
    \end{split}
    \label{kjh}
    \end{equation}
where the stress-energy tensor $T_{\mu\nu}=T_{\mu\nu}^{mat}+T_{\mu\nu}^{B}$, with  $T_{\mu\nu}^{mat}$ is defined analogous to Eq.(\ref{matt}) and 
\begin{equation}
    T_{\mu\nu}^{B}=B_{\mu\sigma}B_{\nu}^{\,\,\sigma}-\frac{1}{4}g_{\mu\nu}B^{\alpha}_{\,\,\sigma}B^{\sigma}_{\,\,\alpha}-Vg_{\mu\nu}+2V^{\prime}B_{\mu}B_{\nu}.
    \end{equation}

In order to make a simplification, we project Eq.(\ref{kjh}) along $g^{\mu\nu}$, $B^{\mu}$ and $B^{\mu}B^{\nu}$, respectively, to obtain the following expressions
\begin{eqnarray}
R(\Gamma)&=&-\kappa^2 T;\label{RT}\\
    \nonumber B^{\mu} R_{\mu\nu}(\Gamma) &=& \frac{4 \kappa^{2}}{(4+3\xi B^{2})}\bigg\{ T_{\mu\nu} B^{\mu}-\frac{B_{\nu}T}{2} - \frac{2\xi B_{\nu}}{4 + 5\xi B^{2}}\bigg[ B^{\alpha}B^{\beta}T_{\alpha\beta} - \\
    &-&\frac{1}{4}B^{2} T \left( 1 - \frac{3}{4}\xi B^{2} \right)    \bigg]      \bigg\};\label{b1}\\
    B^{\mu}B^{\nu} R_{\mu\nu}(\Gamma) &=& \frac{4\kappa^{2}}{4+5\xi B^{2}} \left[ B^{\mu}B^{\nu}T_{\mu\nu} -\frac{B^{2}T}{8} (4 + \xi B^{2})        \right]. \label{b2}
\end{eqnarray}
Plugging Eqs.(\ref{RT}), (\ref{b1}), (\ref{b2}) into Eq.(\ref{kjh}), we have
\begin{equation}
\begin{split}
    R_{\mu\nu}(h)&=\kappa^{2}_{eff}\bigg\{T_{\mu\nu}-\frac{1}{2}g_{\mu\nu}T+\frac{2\xi g_{\mu\nu}}{(4+5\xi B^{2})} \left[ B^{\alpha}B^{\beta}T_{\alpha\beta} -\frac{B^{2}T}{16} (4 - 3\xi B^{2})        \right]+\\
    &+\frac{8\xi}{(4+3\xi B^2)}B_{(\mu}\left[T_{\nu)\alpha}B^{\alpha}-\frac{B_{\nu)}T}{2}-\frac{2\xi B_{\nu)}}{(4+5\xi B^2)}\left(B^{\alpha}B^{\beta}T_{\alpha\beta}-\frac{1}{4}B^{2}T\left(1-\frac{3}{4}\xi B^2\right)\right)\right]\bigg\},
    \label{RR}
    \end{split}
\end{equation}
where $\kappa^{2}_{eff}=\dfrac{\kappa^{2}}{1-\frac{\xi B^2}{4}}$.

Recalling that the connection is given by the Christoffel symbols of $h_{\mu\nu}$. In addition, using Eqs.(\ref{metric1}) and (\ref{metric2}), one can rewrite Eq.(\ref{RR}) in a dynamical field equation for the auxiliary metric $h_{\mu\nu}$ in Einstein--like shape. We shall display their explicit form later.  

Let us now turn our attention to the bumblebee field equation. Varying the action (\ref{bumb}) with respect to $B_{\mu}$, one finds
\begin{equation}
    \nabla^{(g)}_{\mu}B^{\mu\alpha}=-\frac{\xi}{\kappa^2}g^{\nu\alpha}B^{\mu}R_{\mu\nu}(\Gamma)+\frac{\xi}{4\kappa^2}B^{\alpha}R(\Gamma)+2V^{\prime}(B^{\mu}B_{\mu}\pm b^2)B^{\alpha},
    \label{bumb2}
\end{equation}
where the prime above stands for the derivative with respect to the argument of the potential $V$, and $\nabla^{(g)}_{\mu}$ is the covariant derivative defined in terms of the Levi--Civita connection of $g_{\mu\nu}$. Inserting Eqs.(\ref{RT}) and (\ref{b1}) into Eq.(\ref{bumb2}), one gets a Proca--like equation
\begin{eqnarray}
\nabla^{(g)}_{\mu}B^{\mu\alpha}=\mathcal{M}^{\alpha}_{\,\,\,\nu}B^{\nu},
\label{proca}
\end{eqnarray}
where we have defined the effective mass--squared tensor by
\begin{eqnarray}
\nonumber \mathcal{M}^{\alpha}_{\,\,\,\nu}&=&\bigg\{2V^{\prime}+\frac{\xi T\left(4-3\xi B^2\right)}{4\left(4+3\xi B^2\right)}+\frac{8\xi^2}{\left(4+3\xi B^2\right)\left(4+5\xi B^2\right)}\bigg[B^{\mu}B^{\lambda}T_{\mu\lambda}-\\
&-&\frac{1}{4}B^2 T\left(1-\frac{3}{4}\xi B^2 \right)\bigg]\bigg\}\delta^{\alpha}_{\,\,\,\nu}-\frac{4\xi}{\left(4+3\xi B^2\right)}T^{\alpha}_{\,\,\,\nu}.
\label{Mat}
\end{eqnarray}
Note that the new unconventional interaction terms between the bumblebee field and the stress--energy tensor allow us to have new effects in a distinguishing way to the metric case. One can cite, for example, the mechanism of spontaneous vectorization that occurs when the bumblebee field spontaneously acquires an effective mass near high-density compact objects \cite{Ramazanoglu:2017xbl, Ramazanoglu:2019jrr, Cardoso:2020cwo}. More so, due to the negative sign between the first and second terms in Eq.(\ref{Mat}), the determinant of the effective mass-squared matrix can assume negative values leading to tachyonic--like instabilities. 

 Observe, however, that Eq.(\ref{proca}) can be cast into a more convenient form by introducing a conserved current, $J^{\mu}$. To see that in more detail, let us take the divergence of Eq.(\ref{proca}), and then one obtains
\begin{equation}
    \nabla^{(g)}_{\mu}J^{\mu}=0,
    \label{fg}
\end{equation}
where 
\begin{equation}
    J^{\mu}=\mathcal{M}^{\mu}_{\,\,\,\nu}B^{\nu}.
\end{equation}
By defining the bumblebee field equation in terms of the conservation of a current, Eq.(\ref{fg}), it permits us to find  regular solutions easier. We shall discuss on exact solutions of the metric--affine bumblebee model in the next section.

\section{Application: A static and spherically symmetric solution in metric-affine traceless bumblebee model}
\label{Application}

This section is aimed at providing a static and spherically symmetric solution for the metric-affine traceless bumblebee model discussed before. Initially, let us restrict our attention to vacuum solutions which are featured by the  absence of matter sources, $T_{\mu\nu}^{(mat)}=0$. Apart from that, we fix the bumblebee field to assume its vacuum expectation value, i.e., $<B_{\mu}>=b_{\mu}$, that leads to $V=0$ and $V^{\prime}=0$.

In this scenario, we shall start with the field equations displayed in the last subsection. The first important ingredient is the metric. Notice that Eq.(\ref{RR}) is the dynamical equation for the metric  $h_{\mu\nu}$. Thereby, it is more convenient to manipulate the field equations in the Einstein frame. Since we are interested in static and spherically symmetric solutions, then the line element in spherical coordinates $(t,r,\theta,\phi)$ takes the following form:
\begin{equation}
    ds^2_{(h)}=-e^{2\sigma(r)}dt^2+e^{-2\rho(r)}dr^2+r^2\left(d\theta^2 +\sin^{2}{\theta}d\phi^2\right),
\end{equation}
where $\sigma(r)$ and $\rho(r)$ are the metric functions. The second ingredient is the form of $b_{\mu}$. In order to find a regular solution, we impose that the norm of the conserved current in the Einstein frame, $J^{2}=h^{\mu\nu}J_{\mu}J_{\nu}$, vanishes throughout the space-time, which guarantees that the current does not diverge at the horizon \footnote{A similar choice has been chosen in the context of Galileons in \cite{Rinaldi, Babichev}.}. Such a requirement is fulfilled assuming $b_{\mu}$ to have the form:
\begin{equation}
    b_{\mu}=[0,b(r),0,0],
    \label{bumbp}
\end{equation}
 which leads to the vanishing of the field strength associated with it, $b_{\mu\nu}=\left(db\right)_{\mu\nu}$. As a consequence, $T_{\mu\nu}$ and also $J_{\mu}$ vanish even without imposing any previous condition on $b(r)$. Before proceeding further, it is worth calling attention to the conventions that we shall adopt here: tilded objects are defined in the Einstein frame, namely, an index can be risen or lowered using the auxiliary metric, $h^{\mu\nu}$. For example, $\tilde{b}^{\mu}=h^{\mu\nu}b_{\nu}$. Note that although $b^2=g^{\mu\nu}b_{\mu}b_{\nu}$ possesses an explicit dependence of $g^{\mu\nu}$, one can define a new object $\tilde{b}^2=h^{\mu\nu}b_{\mu}b_{\nu}$, which depends on $h_{\mu\nu}$. Both of them are algebraically related to each other by $\tilde{b}^{2}=b^2\dfrac{\left(1+\frac{3\xi b^2}{4}\right)^{1/2}}{\left(1-\frac{\xi b^2}{4}\right)^{3/2}}$. Thereby, $b^2$ can properly be written in terms of $\tilde{b}^2$.  Furthermore, as we mentioned before, $b^2$ is a real constant, as $\tilde{b}^{2}$ is.

 The requirement that $\tilde{b}^{2}=const.$ leads to $b(r)=|\tilde{b}|e^{-\rho(r)}$. Putting all the aforementioned features together,  the vacuum field equations in the Einstein frame (\ref{RR}) are drastically simplified to
 \begin{equation}
     R_{\mu\nu}(h)=0,
 \end{equation}
whose solution is the well--known Schwarzschild line element
\begin{equation}
    ds^{2}_{(h)}=-\left(1-\frac{2M}{r}\right)dt^2 +\frac{dr^2}{\left(1-\frac{2M}{r}\right)} +r^{2}\left(d\theta^2 +\sin^{2}{\theta} d\phi^2\right),
    \label{hm}
\end{equation}
and, the VEV, is given by
\begin{equation}
    b_{\mu}=\left[0,\frac{|\tilde{b}|}{\sqrt{1-\frac{2M}{r}}},0,0\right].
\end{equation}
 Notice that although $b_{\mu}$ diverges at the horizon -- probably, due to an effect of a ``bad'' gauge choice -- the physical observables are characterized by the scalar invariants built up from $b_{\mu}$ which are finite at the horizon. For example, $b^2=const$, and $J^2=0$, by construction, and $b^{\mu}b^{\nu}R_{\mu\nu}=0$.

In order to find the metric $g_{\mu\nu}$, we substitute Eq.(\ref{hm}) in Eq.(\ref{metric2}), identifying $B_{\mu}=b_{\mu}$. After that, one obtains the line element for $g_{\mu\nu}$, namely,
\begin{equation}
    ds^2_{(g)}=-\frac{\left(1-\frac{2M}{r}\right)}{\sqrt{\left(1+\frac{3X}{4}\right)\left(1-\frac{X}{4}\right)}}dt^2+\frac{dr^2}{\left(1-\frac{2M}{r}\right)}\sqrt{\frac{\left(1+\frac{3X}{4}\right)}{\left(1-\frac{X}{4}\right)^3}}+r^{2}\left(d\theta^2 +\sin^{2}{\theta}d\phi^2\right),
    \label{metric3}
\end{equation}
where we have used the shorthand notation: $X=\xi b^2$, which  effectively represents the Lorentz-violating coefficient. Note that the line element in Eq.(\ref{metric3}) describes a LSB modified Schwarzschild metric. The LSB coefficient affects not only the radial component of the metric $g_{\mu\nu}$, but also its temporal component in a distinguishing way. In order to investigate the solution, let us compute the Kretschmann scalar invariant:
\begin{eqnarray}
\nonumber K=R_{\alpha\beta\mu\nu}R^{\alpha\beta\mu\nu}&=&\frac{1}{r^{6}\left(4+3X\right)^{3/2}}\Bigg[48\,XMr\sqrt {4+3\,X}+32\,MXr\sqrt {4-X}-\\
\nonumber&-&12\,M{X}^{2}r\sqrt {4-X}+32
\,{r}^{2}\sqrt {4+3\,X}+192\,{M}^{2}\sqrt {4+3\,X}-\\
\nonumber&-&32\,{r}
^{2}\sqrt {4-X}-16\,{r}^{2}X\sqrt {4-X}- 12\,{X}^{2}Mr\sqrt {4+3\,X}+\\
\nonumber &+&6\,{r}^{2}{X}^{2}\sqrt {4-X}+64\,Mr\sqrt {4-X}-144\,X{M}^{2}\sqrt {4+3\,X}-\\
\nonumber&-&3\,{M
}^{2}{X}^{3}\sqrt {4+3\,X}+36\,{M}^{2}{X}^{2}\sqrt {4+3\,X}+3\,{X}^{2}
{r}^{2}\sqrt {4+3\,X}+\\
&+&{X}^{3}Mr\sqrt {4+3\,X}-64\,Mr\sqrt {4+3\,X}-\frac{1}{4}\,{X}^{3}{r}^{2}\sqrt {4+3\,X}\Bigg].
\end{eqnarray}
Clearly, the scalar invariant shows that the effects of LSB (expressed through $X$) cannot be absorbed by a simple coordinate rescaling. If we consider $X=0$, we recover the usual result for the Schwarzschild metric, $K_{S}=\frac{48 M^2}{r^6}$, as expected. Since the coefficient for Lorentz violation  must be suppressed by a typical high energy scale \cite{KosLi}, then, assuming $X<<1$, the effects of LSB at the first-order expansion in $X$ of the Kretschmann invariant read:
\begin{eqnarray}
K=K_{S}+\frac{12MX}{r^6}\left(r-6M\right)+\mathcal{O}(X^2).
\end{eqnarray}
Note that the structure of singularities holds the same in comparison to the Schwarzschild one. At $r=6M$, the leading term vanishes, so that the effects of LSB at the first-order level are eliminated, although it cannot be directly realized from the metric in Eq. (\ref{metric3}).

\subsection{Geodesics}

Having the knowledge of the LSB metric  (\ref{metric3}), our focus now is obtaining information about the effects of LBS in the geodesic trajectories of particles moving in this spacetime. Since we are dealing with a static and spherically symmetric metric,  there are two Killing vectors associated with it, namely: $\partial_{t}$ and $\partial_{\phi}$. As a result, it suffices to aim at the radial geodesics. In order to find the geodesics of point particles, we begin with the following Lagrangian \cite{Wald}
\begin{equation}
    \mathcal{L}=g_{\mu\nu}\dot{x}^{\mu}\dot{x}^{\nu},
\end{equation}
where $\mathcal{L}$ can take the values: $-1,0,1$, corresponding to the timelike, null and spacelike geodesics, respectively. The dot in the former equation means that there is a derivative with respect to an affine parameter denoted by $\lambda$. Then, the velocity is defined by $\dot{x}^{\mu}\equiv \frac{dx^{\mu}}{d\lambda}$. As it is well known, the motion of particles is independent of the angular coordinate $\theta$; so, for the sake of simplicity, we restrict the particle to move in the equatorial plane, $\theta=\frac{\pi}{2}$. In this case, for the metric (\ref{metric3}), we have
\begin{equation}
    \mathcal{L}=-\frac{\left(1-\frac{2M}{r}\right)}{\sqrt{\left(1+\frac{3X}{4}\right)\left(1-\frac{X}{4}\right)}}\dot{t}^2+\frac{1}{\left(1-\frac{2M}{r}\right)}\sqrt{\frac{\left(1+\frac{3X}{4}\right)}{\left(1-\frac{X}{4}\right)^3}}\,\dot{r}^{2}+r^{2}\dot{\phi}^2.
    \label{geo}
\end{equation}
Here, there are two conserved quantities $E$ and $L$, which are explicitly found by using the Euler--Lagrange equations for the coordinates $t$ and $\phi$, respectively. By doing so, we obtain
\begin{eqnarray}
\label{E1}E&=&\frac{\left(1-\frac{2M}{r}\right)}{\sqrt{\left(1+\frac{3X}{4}\right)\left(1-\frac{X}{4}\right)}}\dot{t},\\
\label{L1}L&=&r^2 \dot{\phi}.
\end{eqnarray}
Putting Eqs.(\ref{E1}), (\ref{L1}) into Eq.(\ref{geo}), we get the radial geodesic equation 
 \begin{equation}
     \frac{\dot{r}^2}{2}=\left(1-\frac{X}{4}\right)^{2}\frac{E^2}{2}-\frac{1}{2}\left(1-\frac{2M}{r}\right)\sqrt{\frac{\left(1-\frac{X}{4}\right)^3}{1+\frac{3X}{4}}}\left(\frac{L^2}{r^2}-\mathcal{L}\right),
     \label{13}
 \end{equation}
which can be rewritten like
\begin{equation}
    \frac{\dot{r}^2}{2}=\mathcal{E}-V_{eff},
    \label{er}
\end{equation}
 where we have defined the following quantities: 
 \begin{eqnarray}
     \label{me}\mathcal{E}&=&\left(1-\frac{X}{4}\right)^{2}\frac{E^2}{2},\\
     \label{vef}V_{eff}&=&\frac{1}{2}\left(1-\frac{2M}{r}\right)\sqrt{\frac{\left(1-\frac{X}{4}\right)^3}{1+\frac{3X}{4}}}\left(\frac{L^2}{r^2}-\mathcal{L}\right).
\end{eqnarray}
Eq.(\ref{er}) describes the motion of a point particle with a unit mass and total energy $\mathcal{E}$ (\ref{me}) in the presence of the effective potential $V_{eff}$ present in Eq. (\ref{vef}). In what follows, we shall investigate the impact of LSB in the geodesics for massive and massless test particles in the innermost regions, which is supposed to depart from the behavior of GR. 

\subsection{Timelike geodesics and the advance of Mercury's perihelion in the LSB Schwarzschild spacetime}

As one can be seen from Eq.(\ref{er}), the radial motions for massive particles are confined to the regions where $\mathcal{E}-V_{eff}>0$. The turning points occur at $\mathcal{E}=V_{eff}$. Observe that the circular orbits ($r_{0}$) arise when the effective potential is flat, i.e., $\frac{d V_{eff}}{dr}\big|_{r=r_{0}}=0$, and these orbits are stable (SCO) whether $\frac{d^{2} V_{eff}}{dr^{2}}\big|_{r=r_{0}}>0$. In other words, it corresponds to the situation in which a particle tends to return to its radial equilibrium ($r_{0}$), even if it suffers a small displacement from its orbit.   

The next step is to investigate the behavior of the effective potential for massive particles. As discussed before, the effective potential for massive particles, which corresponds to timelike geodesics is achieved by taking  $\mathcal{L}=-1$ in Eq.(\ref{vef}). Note that the zero of the effective potential is located at the horizon, that is, at $r=2M$.    
 
We now investigate how the corrections of LSB affect the innermost stable orbit (ISCO) -- which corresponds to the limiting situation where the two circular orbits approach until they collapse one into another -- compared with the GR results. First of all, it is necessary to solve the equation $\frac{dV_{eff}}{dr}=0$ as a function of $X$. By doing so, however, one concludes that the LSB does not affect ISCO. On the other hand, it is expected that non-circular orbits be affected by LSB. Now, let us illustrate this situation by considering the precession of Mercury's perihelion. As it is well known, the first step is finding the radial coordinate $r$ in terms of the angular ones $\phi$, i.e, $r(\phi)$. To do that, we substitute Eq.(\ref{L1}) in Eq.(\ref{er}), which reads
\begin{equation}
    \left(\frac{dr}{d\phi}\right)^2 = \frac{2\mathcal{E}}{L^2}r^4-\sqrt{\frac{\left(1-\frac{X}{4}\right)^3}{1+\frac{3X}{4}}}\left[\left(\frac{r^4}{L^2}-\frac{2Mr^3}{L^2}\right)\left(\frac{L^2}{r^2}+1\right)\right].
\end{equation}
 Now, for convenience, we introduce the new coordinate $y=\dfrac{L^2}{Mr}$, so that the above expression can be given by
\begin{equation}
    \left(\frac{dy}{d\phi}\right)^2+\sqrt{\frac{\left(1-\frac{X}{4}\right)^3}{1+\frac{3X}{4}}}\left[\frac{L^2}{M^2}+y^2 -2y-\frac{2M^2}{L^2}y^3\right]=\frac{2L^2}{M^2}\mathcal{E}.
    \label{kkk}
\end{equation}
The second standard step is rewriting it as a second-order differential equation. This is succeeded by differentiating Eq.(\ref{kkk}) with respect to $\phi$ and also proceeding with further simple algebraic manipulations. In this way, we obtain
\begin{equation}
    \frac{d^2 y}{d\phi^2}+\sqrt{\frac{\left(1-\frac{X}{4}\right)^3}{1+\frac{3X}{4}}}\left(y-1-\frac{3M^2}{L^2}y^2\right)=0.
    \label{edf}
\end{equation}
In contrast to the metric case \cite{Casana}, the LV corrections do not affect the first term in the former equation. They only impact the term within parentheses instead. When LV coefficients are no longer taken into account, we recover GR as expected. The best manner to observe their effects is by treating Eq.(\ref{edf}) perturbatively. Thus, the solution can be set into the form
 \begin{equation}
     y=y_{0}+y_{1}+...,
     \label{pert}
 \end{equation}
where $y_{0}$ is the unperturbed case, $y_{1}$ is the first--order perturbed solution. It is worthy to be mentioned that higher--order corrections will be neglected in the perturbative scheme adopted here. Plugging Eq.(\ref{pert}) into Eq.(\ref{edf}), we find
 \begin{equation}
     \frac{d^2 y_{0}}{d\phi^2}+\sqrt{\frac{\left(1-\frac{X}{4}\right)^3}{1+\frac{3X}{4}}}\left(y_{0}-1\right)=0,
     \label{pç}
 \end{equation}  
which is the  zeroth--order piece  of the full Eq. (\ref{edf}) and 
\begin{equation}
   \frac{d^2 y_{1}}{d\phi^2} + \sqrt{\frac{\left(1-\frac{X}{4}\right)^3}{1+\frac{3X}{4}}}y_1 =\sqrt{\frac{\left(1-\frac{X}{4}\right)^3}{1+\frac{3X}{4}}}\frac{3M^2}{L^2} y_{0}^2,
   \label{kdl}
\end{equation}
is the first--order part of Eq.(\ref{edf}). The solution for Eq.(\ref{pç}) is given by
\begin{equation}
    y_{0}=1+e\cos{\left(\sqrt{\frac{\left(1-\frac{X}{4}\right)^3}{1+\frac{3X}{4}}}\phi\right)},
    \label{kde}
\end{equation}
where $e$ is the eccentricity of the orbits. Note that this equation resembles the standard Newtonian result. Inserting Eq.(\ref{kde}) into Eq.(\ref{kdl}), we are able to get the solution for the first--order equation, namely,
\begin{equation}
    y_1=\frac{3M^2}{L^2}\left[\left(1+\frac{1}{2}e^2\right) -\frac{1}{6}\cos{\left(2\sqrt{\frac{\left(1-\frac{X}{4}\right)^3}{1+\frac{3X}{4}}}\phi\right)}+e\sqrt{\frac{\left(1-\frac{X}{4}\right)^3}{1+\frac{3X}{4}}}\phi\sin{\left(\sqrt{\frac{\left(1-\frac{X}{4}\right)^3}{1+\frac{3X}{4}}}\phi\right)}\right].
\end{equation}
From the physical viewpoint, only the third term plays an important role in our analysis, since the first one is nothing more than a constant, while the second one oscillates close to zero. Thereby, the solution of Eq. (\ref{pert}), up to first-order, may be cast
\begin{equation}
y=1+e\cos{\left(\sqrt{\frac{\left(1-\frac{X}{4}\right)^3}{1+\frac{3X}{4}}}\phi\right)}+  \frac{3M^2}{L^2}e\sqrt{\frac{\left(1-\frac{X}{4}\right)^3}{1+\frac{3X}{4}}}\phi\sin{\left(\sqrt{\frac{\left(1-\frac{X}{4}\right)^3}{1+\frac{3X}{4}}}\phi\right)}.
\end{equation}
It is observed from the experimental data that the quantity $\epsilon\equiv \frac{3M^2}{L^2}\ll 1$ for Mercury \cite{Dinverno}. Therefore, the above equation up to first-order in $\epsilon$ can be rewritten as
\begin{equation}
    y=1+\cos{\left(\left(1-\epsilon\right)\sqrt{\frac{\left(1-\frac{X}{4}\right)^3}{1+\frac{3X}{4}}}\phi\right)}.
\end{equation} With this, we can define the period of the non--circular orbits being
\begin{equation}
T=\frac{2\pi}{(1-\epsilon)}\left(\frac{\left(1-\frac{X}{4}\right)^3}{1+\frac{3X}{4}}\right)^{-\frac{1}{2}}\approx 2\pi + \Delta\alpha,
\end{equation}
where the angle $\Delta\alpha$ measures the advance of perihelion. Assuming that LV coefficients are so small, we get 
\begin{equation}
    \Delta\alpha=2\pi\epsilon+\frac{3}{2}\pi X=\Delta\alpha_{0}+\delta_{_{LV}}\alpha,
    \label{lo}
\end{equation}
with $\Delta\alpha_{0}$ being the standard contribution from GR, which is usually rewritten as
\begin{equation}
    \Delta\alpha_{0}=2\pi\epsilon=\frac{6\pi GM}{c^2 (1-e^2)a},
\end{equation}
where we have restored Newton's constant $G$ and the speed of light $c$; furthermore, the constant $a$ 
 denotes the semi--major axis of the orbital ellipse. The second contribution on the \textit{r.h.s} of Eq.(\ref{lo}) gives rise to new effects ascribed to LSB. It is important to highlight that such a feature does not have any similarity with GR. Explicitly, we have
 \begin{equation}
     \delta_{_{LV}}\alpha=\frac{3}{2}\pi X.
 \end{equation}
 In other words, this equation accounts for the LV contributions for the advance of Mercury's perihelion in the metric--affine bumblebee gravity. More so, it is bigger than the results currently encountered in the literature to the metric case \cite{Casana}.  

 \subsubsection{Estimation of the LSB coefficient from the advance of Mercury's perihelion}
 Our aim here is to estimate $X$ from astrophysics  data from the advance of Mercury's perihelion. Initially, the theoretical result predicted from GR is $\Delta\alpha_{0}=42.981 \mbox{as}/\mbox{cty}$ \cite{link,Dinverno, Carroll:2004st, Iorio:2018adf}, where the units ``as'' and ``cty'' mean arcseconds and century respectively. By using recent observational data \cite{Pitjev,Pitjev2}, which previews a discrepancy order of $0.002\pm 0.003$ as/cty, we are capable of estimating an upper bound for $X$. In fact,  as long as LSB effects are not sensitive to the current experiments, $\delta_{_{LV}}\alpha$ should be smaller than the observational error ($0.003$ as/cty or, equivalently, $72.3\times 10^{-11}$ as/orbit). In this way, it is straightforward to conclude that LV coefficient must satisfy the upper bound $X<7.4\times 10^{-12}$. This result is more stringent than the ones found in the metric approach \cite{Casana}.

\subsection{Null geodesics and the deflection of light in the LSB Schwarzschild spacetime}

\begin{figure}[!h]
    \centering
    \includegraphics[width=0.50\textwidth]{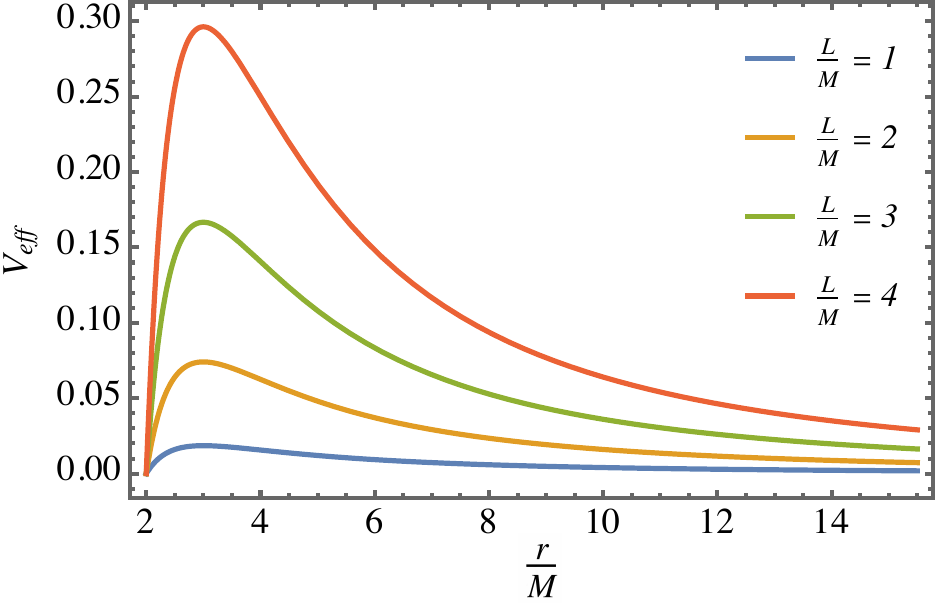}
    \caption{This plot displays the behavior of the effective potential for null geodesics for a variety of values of $\frac{L}{M}$. We take $X=0.01$.}
    \label{fig3}
\end{figure}

Now, we turn our attention to null geodesics which are recovered by taking $\mathcal{L}=0$ in Eq.(\ref{vef}). Note from Fig.(\ref{fig3}) that the shape of the effective potential does not depend on $L$ or $\frac{L}{M}$, as it already happens in GR. 

We shall proceed with the computation of the deflection of light in the weak field approach. Let us begin considering Eq.(\ref{13}) for null geodesics, i.e.,
\begin{equation}
     \dot{r}^2=\left(1-\frac{X}{4}\right)^{2}E^2-\left(1-\frac{2M}{r}\right)\sqrt{\frac{\left(1-\frac{X}{4}\right)^3}{1+\frac{3X}{4}}}\left(\frac{L^2}{r^2}-\mathcal{L}\right).
\end{equation}
We can combine the above equation with those describing the conserved quantities Eqs.(\ref{E1}, \ref{L1}) to find 
\begin{equation}
    \frac{d\phi}{dr}=\left[\left(1-\frac{X}{4}\right)^{2}\frac{r^4}{\beta^2}-\left(1-\frac{2M}{r}\right)\sqrt{\frac{\left(1-\frac{X}{4}\right)^{3}}{\left(1+\frac{3X}{4}\right)}}r^2\right]^{-1/2},\label{fi}
\end{equation}
where we have defined the impact parameter $\beta=\frac{L}{E}$. We shall use this equation to calculate the bending of light in what follows.

Denoting the turning point of a particular orbit by $r_{0}$, we are able to conclude that $\mathcal{E}(r_{0})=V_{eff}(r_{0})$ which implies that 
\begin{equation}
    \frac{1}{\beta^2}=\frac{\left(1-\frac{2M}{r_{0}}\right)}{r_{0}^{2}}\frac{1}{\sqrt{\left(1-\frac{X}{4}\right)\left(1+\frac{3X}{4}\right)}}.
    \label{kl}
\end{equation}
Integrating Eq.(\ref{fi}), we obtain
\begin{equation}
    \Delta\phi=2\int_{r_{0}}^{\infty}\left[\left(1-\frac{X}{4}\right)^{2}\frac{r^4}{\beta^2}-\left(1-\frac{2M}{r}\right)\sqrt{\frac{\left(1-\frac{X}{4}\right)^{3}}{\left(1+\frac{3X}{4}\right)}}r^2\right]^{-1/2} dr,
    \label{kj}
\end{equation}
where the factor $2$ in the above equation arises from the fact that the region before and after the turning point coincides. The former equation can be rewritten in terms of the new variable defined by $u=\frac{1}{r}$, then Eq.(\ref{kj}) becomes
\begin{equation}
    \Delta\phi=2\int_{0}^{u_{0}}\left[\left(1-\frac{X}{4}\right)^{2}\frac{1}{\beta^2}-\left(1-2Mu\right)u^{2}\sqrt{\frac{\left(1-\frac{X}{4}\right)^{3}}{\left(1+\frac{3X}{4}\right)}}\right]^{-1/2} du,
    \label{uu}
\end{equation}
where $u_{0}=\frac{1}{r_{0}}$. Substituting Eq.(\ref{kl}) in Eq.(\ref{uu}), one gets
\begin{equation}
    \Delta\phi=2\sqrt{\frac{\left(1-\frac{X}{4}\right)^{3}}{\left(1+\frac{3X}{4}\right)}}\int_{0}^{u_{0}}\left[\left(1-2GMu_{0}\right)u_{0}^{2}-\left(1-2GMu\right)u^{2}\right]^{-1/2} du,
\end{equation}
where we have restored Newton's constant, $G$, for convenience.
Let us now solve the former integral perturbatively since there is no an exact solution. Therefore, we expand Eq.(\ref{kj}) around $GM=0$ to obtain
\begin{equation}
    \Delta \phi=\sqrt{\frac{\left(1-\frac{X}{4}\right)^{3}}{\left(1+\frac{3X}{4}\right)}}\left(\pi + \frac{4GM}{\beta}+\mathcal{O}\left((GM)^2\right)\right).
\end{equation}
 As discussed before, the coefficient for Lorentz violation must be small, then assuming $X<<1$, we have
\begin{equation}
    \Delta\phi=\pi\left(1-\frac{3X}{4}\right)+\frac{4GM}{\beta}\left(1-\frac{3X}{4}\right)+\mathcal{O}\left((GMX)^2 \right).
\end{equation}
It is well known that the angular deflection in the weak field approximation is defined by \cite{Dinverno}
\begin{equation}
\begin{split}
    \delta\phi&=\Delta\phi-\pi\\
              &=-\frac{3}{4}X\pi+\frac{4GM}{\beta c^2}-\frac{3GMX}{\beta c^2}+\mathcal{O}\left((GMX/c^2)^2 \right),
    \label{klo}
    \end{split}
\end{equation}
where we have restored the speed of light, $c$. Note that the first term in the above equation is the angular deflection due entirely to the LSB which is a negative correction different from the metric bumblebee model \cite{Casana}. The second term in Eq.(\ref{klo}) is the standard GR contribution for the compact object with mass $M$. Ultimately, the third term in Eq.(\ref{klo}) is the second LSB correction for the compact object with mass $M$, such a correction is essentially new since it does not appear in the metric theory \cite{Casana}. Hence, the angular deflection due to LSB reads:
\begin{equation}
    \delta_{_{LV}}\phi=\frac{3}{4}X\pi+\frac{3GMX}{\beta c^2},
    \label{kdel}
\end{equation}
up to first order in $X$ and $\frac{GM}{c^2}$.

\subsubsection{ Estimation of the LSB coefficient from the deflection of light}

In order to estimate the LSB coefficient from recent astronomical data of the deflection of light, we shall consider the parameterized post-Newtonian (PPN) approach \cite{Misner, Will:2018bme}. In particular, we focus on the PPN parameter, $\gamma$, which is used to parameterize the post--Newtonian contributions of space--curvature to the bending of light. The results of GR are recovered for  $\gamma=1$. Within the PPN formalism, the post-Newtonian deflection angle due to a light ray passing through the massive body of mass $M$ at a distance $\beta$ is given by \cite{Will, Klioner}
\begin{equation}
    \delta_{_{PN}}\phi=\frac{1}{2}(1+\gamma)\frac{4GM}{\beta c^2}\frac{\left(1+\cos{\psi}\right)}{2},
\end{equation}
where $\psi$ is the angle between the massive body and the source. Now, let us consider the particular situation that consists of the Sun being chosen to be the deflecting body, i.e., $M=M_\odot$, and we consider a grazing ray that means taking $\beta\approx R_\odot$ and $\psi\approx 0$. In this case, using the constant values from \cite{Butkevich:2022cas}, we have
\begin{equation}
    \delta_{_{PN}}\phi=\frac{1}{2}(1+\gamma)\delta_{_{GR}}\phi,
\end{equation}
with $\delta_{_{GR}}\phi=1.751557143\,\,\mbox{as}$. Note that, as we called attention, $\gamma=1$ recovers GR. 

A variety of observational data of $\gamma$ from experimental tests has been made over time (see \cite{Shapiro, Bertotti, Robertson, Lebach, Lambert, Shapiro2, Fomalont}, and the references therein). One can remark the latest accuracy of the PPN parameter $\gamma$, which has been made BY using the Very Long Baseline Interferometry (VLBI) technique \cite{Titov}. The accuracy obtained was $9\times 10^{-5}$. Here, we assume that the effects of LSB exist in nature and they have not been observed yet by these experiments because their effects are supposed to be suppressed by a high--energy scale. Therefore, the angular deflection due to LSB must be upper bounded by the stringent constraint $\delta_{_{LV}}\phi<0.0788201\, \mbox{mas}$, which leads to $X<1.62\times 10^{-10}$.



\section{Summary and conclusion}
\label{summary}

We dealt with a metric-affine generalization of the gravitational sector of the SME imposing projective invariance. To attain this symmetry, we required some conditions to the LV coefficients, namely: $s^{\mu\nu}$ be symmetric in their two indices and $t^{\alpha\beta\mu\nu}=0$. Under these circumstances, we rigorously obtained the field equations and we found that the solution of the connection equation was simply the Christoffel symbols of a disformally related metric, where the disformal piece was determined by the LV coefficient $s^{\mu\nu}$. As we saw, the action admitted an Einstein--frame representation. 

Having knowledge of the connection, we considered a particular case corresponding to the metric--affine bumblebee gravity in the vacuum. In this context, upon some algebraic manipulations, we were able to rewrite the metric field equation as an Einstein--like vacuum one for the auxiliary metric, $h_{\mu\nu}$. Also, we found an interesting result: the background metric ($g_{\mu\nu}$) picked up contributions stemming from the bumblebee VEV, which was assumed to be a space--like vector. Such a solution described a Schwarzschild--like metric. Remarkably, different from the metric case \cite{Casana}, our solution presented a non--trivial redshift factor induced by the effects of the non--metricity, which was sourced by the LV coefficient, $X=\xi b^2$, as discussed in \cite{Paulo4}. 

We investigated the impact of the LSB in local properties of this background. For instance, the geodesics of particles departed from GR ($X=0$), as larger $X$ was, were not an unexpected behavior. Moreover, we computed the advance of Mercury's perihelion and angular deflection of light within the weak field approximation. In both cases, we obtained corrections stemming from the LSB. We provided an upper bound for the LV coefficient, $X\approx 10^{-12}$, from the observational data of the advance of Mercury's perihelion. In addition,  using the most recent observable data from VLBI for the light-bending, we found that the LV coefficient relied on the following stringent constraint $X<1.62\times 10^{-10}$. 

As a natural continuation of this work, we intend to study the implications of our solutions for the deflection of light within the strong field regime. Also, it would be a promising task examining new solutions, such as rotating black holes.

\section*{Acknowledgments}
\hspace{0.5cm}The authors would like to thank the Conselho Nacional de Desenvolvimento Cient\'{\i}fico e Tecnol\'{o}gico (CNPq) and Fundação de Apoio à Pesquisa do Estado da Paraíba (FAPESQ), project No. 200486/2022-5, for financial support. P. J. Porf\'{\i}rio would like to acknowledge the Brazilian agency CNPq for the financial support, grant No. 307628/2022-1. The work by A. Yu. Petrov. has been partially supported by the CNPq project No. 301562/2019-9. Particularly, A. A. Araújo Filho acknowledges the Facultad de Física - Universitat de València and Gonzalo J. Olmo for the kind hospitality when part of this work was made. Moreover, A. A. Araújo Filho has been partially supported by Conselho Nacional de Desenvolvimento Cient\'{\i}fico e Tecnol\'{o}gico (CNPq) - grants No. 200486/2022-5 and 150891/2023-7 - and CAPES-PRINT (PRINT - PROGRAMA INSTITUCIONAL DE INTERNACIONALIZAÇÃO) - 88887.508184/2020-00.

\end{document}